\documentclass[preprint,showpacs,preprintnumbers,amsmath,amssymb]{revtex4}

\usepackage{graphicx}
\usepackage{dcolumn}
\usepackage{bm}

\begin{document}


\title{Comment on ``Exact three-dimensional wave function and the on-shell $t$ matrix for the sharply cut-off
Coulomb potential: Failure of the standard renormalization factor''}

\author{Yuri V. Popov}
 \affiliation{Nuclear Physics Institute, Moscow State University, Moscow 119991, Russia}
\email{popov@srd.sinp.msu.ru}
\author{Konstantin A. Kouzakov}%
\affiliation{Department of Nuclear Physics and Quantum Theory of Collisions, Faculty of Physics, Moscow State University, Moscow 119991, Russia}
 \email{kouzakov@srd.sinp.msu.ru}

\author{Vladimir L. Shablov}
\affiliation{Obninsk State Technical University for Nuclear Power
Engineering, Obninsk, Kaluga
Region 249040, Russia}%

\date{\today}

\begin{abstract}
The solutions analytically derived by Gl\"ockle \emph{et al.}
[Phys. Rev. C {\bf 79}, 044003 (2009)] for the three-dimensional
wave function and on-shell $t$ matrix in the case of scattering
on a sharply cut-off Coulomb potential appear to be fallacious.
And their renormalization factor lacks mathematical grounds.
\end{abstract}

\maketitle

In a recent paper by Gl\"ockle \emph{et al.}~\cite{gloeckle09},
nonrelativistic scattering of two equally charged particles with
mass $m$ interacting via potential $V(r)=\frac{e^2}{r}\Theta(R-r)$
was considered. The authors argued that they analytically derived
the exact wave function and scattering amplitude for arbitrary
values of a cut-off radius $R$. On this basis they obtained a
renormalization factor which relates the scattering amplitude in
the limit $R\to\infty$ with the physical Coulomb scattering
amplitude. The purpose of this Comment is (i) to show that the
analytical results of Gl\"ockle \emph{et al.}~\cite{gloeckle09}
are erroneous for finite values of $R$ and are mathematically
ungrounded in the limit $R\to\infty$ and (ii) to point out a
different renormalization approach which is free from
uncertainties associated with the cut-off renormalization.

The authors of~\cite{gloeckle09} made an unjustified premise that the solution of the
Lippmann-Schwinger equation for $r<R$ obeys the form
%
\begin{equation}
\label{incorrect} \Psi^{(+)}_R({\vec r})=Ae^{i{\vec p}\cdot{\vec
r}}{_1}F_1(-i\eta,1,i(pr-{\vec p}\cdot{\vec r})),
\end{equation}
where $\eta=\frac{me^2}{2p}$ is a Sommerfeld parameter. The
constant~\footnote{The normalization factor
$\frac{1}{(2\pi)^{3/2}}$ is suppressed throughout.}
\begin{equation}
\label{s-wave} A=\frac{1}{{_1}F_1(-i\eta,1,2ipR)}
\end{equation}
was determined in Ref.~\cite{gloeckle09} by
inserting~(\ref{incorrect}) into the Lippmann-Schwinger equation
\begin{equation}
\Psi^{(+)}_R(\vec r)=e^{i\vec p\vec
r}-\frac{\mu e^2}{2\pi}\int \frac{d^3r'}{r'}
\frac{e^{ip|\vec r-\vec r'|}}{|\vec r-\vec r'|}\Theta(R-r')\Psi^{(+)}_R(\vec
r') \label{LS}
\end{equation}
and solving the latter at $r=0$.
In fact, the
correct form of the solution in the interior region $r<R$ is
\begin{equation}
\label{correct} \Psi^{(+)}_R({\vec r})=\frac{1}{4\pi}\int
d^2\hat{k} \mathcal{A}(\hat{k})e^{ip{\hat k}\cdot{\vec
r}}{_1}F_1(-i\eta,1,i(pr-p{\hat k}\cdot{\vec r})),
\end{equation}
where the function $\mathcal{A}(\hat{k})$ is defined on a unit
sphere. To determine $\mathcal{A}(\hat{k})$ one may employ the usual partial wave
formalism (see, for instance, the textbook~\cite{taylor_book}).
Consider the following expansion in Legendre polynomials:
\begin{eqnarray}
\label{expansion_A}
\mathcal{A}(\hat{k})&=&\sum_{l}(2l+1)A_lP_{l}(\hat{p}\cdot\hat{k}).
\end{eqnarray}
Matching the interior
Lippmann-Schwinger solution and its derivative to the exterior
ones at $r=R$ yields
%
\begin{equation}
A_{l}=\frac{i(pR)^{-2}}{W(\psi_l,h^{(1)}_l)(pR)},\label{partial_coef_res}
\end{equation}
where $h^{(1)}_l$ is a spherical Hankel function of the first kind and
$$
\psi_l(pr)=
e^{i\sigma_l}\frac{|\Gamma(l+1+i\eta)|}{\Gamma(1+i\eta)}\frac{(2pr)^l}{(2l+1)!}e^{-ipr}
{_1}F_1(l+1-i\eta,2l+2,2ipr),\nonumber
$$
with the Coulomb phase shift
$\sigma_l=\text{arg}\Gamma(l+1+i\eta)$. It can be checked that $A_0=A$ but $A_{l\geq1}\neq A$, i.e.~(\ref{incorrect}) is clearly invalid.

The expression for
the scattering amplitude (the on-shell $t$ matrix) in
Ref.~\cite{gloeckle09} is invalid as well, since it derives from
the wave function~(\ref{incorrect}).
The limit of vanishing screening ($R\to\infty$) has been
considered previously in the literature (see, for
instance,~\cite{taylor_book,ford64,taylor74} and references
therein). Using
asymptotic forms of $\psi_l$ and $h_l^{(1)}$~\cite{abramowitz} one
readily arrives at
\begin{equation}
A_l\simeq
e^{-\frac{\pi\eta}{2}}\Gamma(1+i\eta)e^{-i\eta\ln(2pR)}+O\left(\frac{1}{pR}\right),\label{limit_coeff}
\end{equation}
provided $l\ll pR$. When $l\gg pR$, the phase shifts behave as
$\delta_l\to0$ due to the angular momentum barrier. The
intermediate situation $l\sim pR$ is very hard to
handle~\cite{ford64}. Thus, the convergence $A_l\to
e^{-\frac{\pi\eta}{2}}\Gamma(1+i\eta)e^{-i\eta\ln(2pR)}$ is not
uniform, i.e. it depends on $l$, and therefore taking the limit
$R\to\infty$ in~(\ref{expansion_A}) presents quite a challenge.
Nevertheless, it can be shown that the asymptotic form for the
scattering amplitude is~\cite{taylor_book,taylor74}
\begin{equation}
\label{ampl_taylor} f_R=e^{-2i\eta\ln
(2pR)}f_c+f_{osc},
\end{equation}
where $f_c$ is the physical Coulomb scattering amplitude. The
first term in the right-hand side of~(\ref{ampl_taylor}) appears
because of~(\ref{limit_coeff}). The term $f_{osc}$ oscillates
rapidly like $\cos(qR)$, where $q$ is the momentum transfer. It
integrates out to zero with the incident wave packet and, hence,
makes no contribution to the cross section as measured in typical
experiments~(see \cite{taylor_book} for details).

The amplitude derived in Ref.~\cite{gloeckle09} in the limit
$R\to\infty$ resembles the form~(\ref{ampl_taylor}), however
its derivation lacks mathematical grounds because it is
carried out using~(\ref{incorrect}) instead of the exact
wave function~(\ref{correct}). The wave function~(\ref{incorrect})
can be presented as a product $C_R\Psi_c^{(+)}$, where $\Psi_c^{(+)}$
is a Coulomb wave and $C_R$ is a constant ($C_{R\to\infty}\to e^{-i\eta\ln(2pR)}$).
The Coulomb wave satisfies a homogeneous Lippmann-Schwinger equation~\cite{west}
\begin{equation}
\Psi^{(+)}_c(\vec r)=-\frac{\mu e^2}{2\pi}\int \frac{d^3r'}{r'}
\frac{e^{ip|\vec r-\vec r'|}}{|\vec r-\vec r'|}\Psi^{(+)}_c(\vec
r'). \label{LS_homo}
\end{equation}
Let us introduce an auxiliary function which is a difference
between the exact wave function~(\ref{correct}) and the wave
function~(\ref{incorrect}) in the limit $R\to\infty$:
\begin{equation}
\label{auxiliary}
\psi_R(\vec{r})=\Psi^{(+)}_R(\vec r)-e^{-i\eta\ln
(2pR)}\Psi^{(+)}_c(\vec r).
\end{equation}
According to~(\ref{LS}) and~(\ref{LS_homo}),
this function satisfies the following equation ($r<R$):
\begin{equation}
\label{LS_aux}
\psi_R(\vec r)=
\psi_R^{(0)}(\vec r)-\frac{\mu e^2}{2\pi}\int
\frac{d^3r'}{r'} \frac{e^{ip|\vec r-\vec r'|}}{|\vec r-\vec
r'|}\Theta(R-r')\psi_R(\vec r'),
\end{equation}
with the inhomogeneous term
\begin{equation}
\psi_R^{(0)}(\vec r)=e^{i\vec p\vec
r}+\frac{\mu e^2}{2\pi}e^{-i\eta\ln(2pR)}\int
\frac{d^3r'}{r'} \frac{e^{ip|\vec r-\vec r'|}}{|\vec r-\vec
r'|}\Theta(r'-R)\Psi^{(+)}_c(\vec r'). \label{inhom}
\end{equation}
For $r\ll R$ one has approximately
$$
\frac{e^{ip|\vec r-\vec r'|}}{|\vec r-\vec r'|}\approx
\frac{e^{ipr'}}{r'}e^{-i\vec p'\vec r}, \qquad \vec p'=\frac{p\vec r'}{r'},
$$
and it can be shown that $\psi_R^{(0)}(\vec r)\approx0$. However this
does not imply that $\psi_R^{(0)}(\vec r)\approx0$ for any $r<R$.
In fact, this is definitely not the case when $r\lesssim R$ (within
the partial wave formalism this situation corresponds to the $l\lesssim pR$ terms).

Using~(\ref{auxiliary}), the scattering amplitude can be presented as
\begin{equation}
\label{amplitude}
f_R=-\frac{\mu e^2}{2\pi}e^{-i\eta\ln(2pR)}\int\frac{d^3r'}{r'} e^{-i\vec p'\vec
r'}\Theta(R-r')\Psi^{(+)}_c(\vec r')-\frac{\mu e^2}{2\pi}\int
\frac{d^3r'}{r'} e^{-i\vec p'\vec r'}\Theta(R-r')\psi_R(\vec r'),
\end{equation}
where $\vec{p}'=p\hat{r}$. Gl\"ockle \emph{et
al.}~\cite{gloeckle09} have studied asymptotic behavior of the
first term only. They unjustifiably have neglected the second term
which, due to nontrivial properties of $\psi_R$ in the region
$r\lesssim R$, potentially can provide a nonvanishing contribution
to the scattering amplitude in the limit $R\to\infty$. Thus, their
analysis is incomplete and the validity of their 
renormalization factor is questionable. In this connection, the results of the numerical calculations
presented in Ref.~\cite{gloeckle09} can not be a decisive
argument in favour of the renormalization factor, for in this particular case one deals with
divergent and rapidly oscillating quantities.

Finally, it is useful to note that the renormalization treatments
involving cut-off Coulomb potentials are of doubtful value from a
practical viewpoint, especially in the case of many-body Coulomb
scattering. In this respect, the methods based on regularization
and renormalization of the Lippmann-Schwinger equations in the
on-shell limit are more efficient. The two-particle case is fully
explored: (i) the Green's function is derived analytically both
in coordinate and in momentum representations~\cite{schwinger},
(ii) an off-shell amplitude is known~\cite{shastry1}, and (iii)
the rules for taking the on-shell limit are formulated~\cite{popov81}. This allows to generalize the two-particle results
to the many-particle case (see, for example,~\cite{shablov_pra}).

\begin{acknowledgments}
We wish to thank Sergue I. Vinitsky for drawing our attention to
the paper by Gl\"ockle \emph{et al.}~\cite{gloeckle09} and to Akram Mukhamedzhanov for useful discussions.
\end{acknowledgments}

\end{document}